\documentclass{article}

    \PassOptionsToPackage{numbers, compress}{natbib}

\usepackage[preprint]{tackling_climate_workshop_style}

\usepackage[utf8]{inputenc} 
\usepackage[T1]{fontenc}    
\usepackage{hyperref}       
\usepackage{url}            
\usepackage{booktabs}       
\usepackage{amsfonts}       
\usepackage{nicefrac}       
\usepackage{microtype}      

\usepackage{sidecap}
\usepackage{graphicx}
\usepackage{wrapfig}      
\sidecaptionvpos{figure}{c}
\usepackage{chemformula} 

\usepackage{xcolor}
\usepackage{ifthen}

\newboolean{review}  
\setboolean{review}{false} 

\definecolor{color1}{HTML}{e41a1c}  
\definecolor{color2}{HTML}{377eb8}      
\definecolor{color3}{HTML}{4daf4a}   
\definecolor{color4}{HTML}{984ea3}
\definecolor{color5}{HTML}{ff7f00}

\newcommand{\Arthur}[1]{\ifthenelse{\boolean{review}}{\textcolor{color1}{Arthur: #1}}{}}
\newcommand{\Mike}[1]{\ifthenelse{\boolean{review}}{\textcolor{color2}{Mike: #1}}{}}
\newcommand{\Ricardo}[1]{\ifthenelse{\boolean{review}}{\textcolor{color3}{Ricardo: #1}}{}}
\newcommand{\Sylvia}[1]{\ifthenelse{\boolean{review}}{\textcolor{color4}{Sylvia: #1}}{}}
\newcommand{\Umangi}[1]{\ifthenelse{\boolean{review}}{\textcolor{color5}{Umangi: #1}}{}}

\title{Detangling the role of climate in vegetation productivity with an explainable convolutional neural network}

\author{%
  Ricardo Barros Louren\c{c}o \\
  School of Earth, Environment \& Society\\
  McMaster University\\
  Hamilton, Ontario, L8S 4L8, Canada \\
  \texttt{barroslr@mcmaster.ca} \\
  \And
  Michael J. Smith \\
  Aspia Space Ltd. \\
  G-50 Tremough Innovation Centre \\
  Penryn, Cornwall, TR10 9TA, UK \\
  \texttt{mike.smith@aspiaspace.com} \\
  \And
  Sylvia Smullin \\
  Figured Eight Consulting, LLC. \\
  Somerville, MA, 02143, USA \\
  \texttt{figured8consulting@gmail.com} \\
  \And
  Umangi Jain \\
  University of Toronto \\
  Toronto, Ontario, M5S 2Z9, Canada \\
  \texttt{umangi.jain@mail.utoronto.ca} \\
  \And
  Alemu Gonsamo \\
  School of Earth, Environment \& Society\\
  McMaster University\\
  Hamilton, Ontario, L8S 4L8, Canada \\
  \texttt{gonsamoa@mcmaster.ca} \\
  \And
  Arthur Ouaknine \\
  McGill University \\ Mila, Quebec AI Institute \\
  Montr\'eal, Qu\'ebec, H2S 3H1, Canada \\
  \texttt{arthur.ouaknine@mila.quebec} \\
}

\begin{document}

\maketitle

\begin{abstract}
Forests of the Earth are a vital carbon sink while providing an essential habitat for biodiversity. Vegetation productivity (VP) is a critical indicator of carbon uptake in the atmosphere. The leaf area index is a crucial vegetation index used in VP estimation. This work proposes to predict the leaf area index (LAI) using climate variables to better understand future productivity dynamics; our approach leverages the capacities of the V-Net architecture for spatiotemporal LAI prediction. Preliminary results are well-aligned with established quality standards of LAI products estimated from Earth observation data. We hope that this work serves as a robust foundation for subsequent research endeavours, particularly for the incorporation of prediction attribution methodologies, which hold promise for elucidating the underlying climate change drivers of global vegetation productivity.
\end{abstract}

\section{Introduction}
\label{sec:intro}
Climate change alters vegetation productivity \citep{Zhu_2016, Gonsamo_Chen_Lombardozzi_2016, Piao_2019} with significant implications for the Earth’s climatic and biological systems. In particular, elevated temperatures have contributed to advanced events such as changing the growing season of plants \citep{Gonsamo_Chen_Ooi_2018} and delaying leaf unfolding and leaf senescence \citep{Beil_Kreyling_Meyer_Lemcke_Malyshev_2021, Geng_2020, Gonsamo_Chen_Ooi_2018}. These impacts, especially apparent in high-latitude countries, indicate changes in ecosystem functioning \citep{Hegland_2009, Parmesan_2006, Parmesan_Yohe_2003, Walther_2002}. It results in various feedback mechanisms affecting the Earth's physical systems such as changing surface reflectivity (albedo) and energy balance \citep{Peñuelas_2009,Richardson_2013,Zeng_2017} while also inducing local scale disturbances in plant-pollinator interactions \cite{Hegland_2009}.

Studies on VP are related to phenology, the former focusing on the estimation of global vegetation production \citep{Piao_2019}, while the latter on the determination of stages of plant development \citep{Piao_Liu_Chen_Janssens_Fu_Dai_Liu_Lian_Shen_Zhu_2019}. For instance, the total rate of carbon photosynthesized by plants (over time, in an area) is characterized as gross primary productivity (GPP) \citep{2023Kooistra}, a form of VP that is a critical indicator of carbon uptake of the atmosphere. Remote sensing observations are widely used to create vegetation indexes (VI) products, such as the normalized difference vegetation index (NDVI), fraction of absorbed photosynthetically active radiation (\textit{f}APAR), and the leaf area index (LAI); the latter being one of the most important products used to estimate global GPP \citep{Piao_2019, 2023Kooistra}. 

This study proposes to predict LAI using climate variables to better estimate and understand the impact of climate change on vegetation productivity worldwide. Since climate factors directly influence plant productivity \cite{Piao_2019}, it is still not yet clear how their dynamics affects vegetation health and growth (VP), and thus the LAI index \cite{Zhu_2016, Piao_2019, Myers-Smith_2020, Gonsamo_2021}.

Successfully predicting LAI using climate variables would open up opportunities for investigating relationships between climate trends and anomalies with vegetation productivity. As a consequence, it will motivate experiments to better quantify the effects of climate forcings on carbon uptake \cite{Piao_2019}. 

These same shifts on vegetation dynamics have a significant impact on global albedo \cite{1984Dickinson}. Therefore, this work could also drive the characterization and quantification of the bidirectional relationships between climate and LAI for many applications other than VP, for instance, vegetation growth causing water depletion in soil, which in turn increases local temperature through surface heating \citep{2022ZHANG,2022Zhu}.

The related work on vegetation productivity applications will be detailed in Section \ref{sec:related_work}. The dataset used and the proposed method for LAI prediction will be presented in Section \ref{sec:data_method}. Finally, Section \ref{sec:results} will detail preliminary results and future research opportunities.

\section{Related work}
\label{sec:related_work}

Mapping and understanding the relationship between vegetation dynamics and climate change has long been a well-established research challenge in global climate change communities.
This is often done through ground measurements \citep{Crimmins_Crimmins_2008, Sonnentag_2012, Richardson_2018, Teeri_Raven_2002, Kampe_2010} or process-based models \citep{Buck-Sorlin2013, Luo_Smith_2022}. The challenge still remains since no existing ground measurements are representative enough of global vegetation with long time-series measurements \citep{Piao_Liu_Chen_Janssens_Fu_Dai_Liu_Lian_Shen_Zhu_2019, Piao_2019, 2023Kooistra}. 
On the other hand, process-based models are too simplistic to reproduce complex global observed vegetation changes, and, at the same time, detangle the correlated roles of climate and other global change agents on LAI dynamics.

\section{Data and method} \label{sec_data}
\label{sec:data_method}

\paragraph{Dataset}
Earth observation data have been used in the format of ground-validated gridded products. The proposed target to predict is the global inventory monitoring and modeling system -- third generation (GIMMS 3g) LAI product \cite{2013Zhu}. The following climate covariates have been considered as input features to tackle the proposed task: cloud cover, precipitation rate, air temperature, and frequency of wet days from the climatic research unit gridded time series (CRU-TS) dataset \cite{Harris_Osborn_Jones_Lister_2020}; incoming solar radiation from the atmospheric forcing data for the community land model (CRUNCEP) dataset \cite{cisl_rda_ds314.3}; albedo from global land surface satellite (GLASS) \cite{2021GLASS}; and soil moisture from global land evaporation Amsterdam model (GLEAM) \cite{2017GLEAM}. The dataset has been formatted to a monthly, half-degree global coverage from January 1982 to December 2015. It has been aggregated, when necessary, and saved in a netCDF format compliant with an Xarray \cite{hoyer2017xarray} dataset structure \cite{2023Rbl}. The test set is defined as the most recent 16 months of the time series and the training set contains the remaining 380 months.

\paragraph{Method}

The proposed experiment integrates a V-Net \citep{2016Milletari} taking as input a 4D array with latitude, longitude, and time axes, for every channel. This way, both spatial and temporal representations are learned using 3D convolutions while preserving information between the encoder and the decoder with its skip connections.

The model is trained using the past climate covariate time series to predict the LAI while reconstructing and predicting the future climate covariates. The input features are shifted on the time axis, using a lag of one month, so that the model reconstructs a climate covariate set of 15 months while predicting one month ahead in the future. 

The training process includes batch of 4 tensors, each one of size $360 \times 720 \times 16 \times 7$, including the entire globe ($360 \times 720$), as a time series of 16 months, with all seven climate covariates. These tensors are randomly built over the training set considering the 380-month period, separated from the 16-month test set. The model predicts a tensor of size $360 \times 720 \times 16 \times 8$,
including a climate covariate reconstruction of 15 months, a one month ahead prediction of these covariates and a LAI prediction of the considered 16 months. It aims at learning spatiotemporal correlations between climate covariates and LAI while having a robust reconstruction of climate phenomena.

The data are preprocessed with min-max normalization, and empty values (derived from Xarray data masking) have been replaced with -1 values. The proposed model has been trained for 73000 iterations to minimize a Huber loss function \citep{1964Huber} using the Adam optimizer \citep{2015KingmaBa}. Our code used for this experiment and the pretrained model are publicly available\footnote{Links will be provided if the article is accepted.}.

\section{Preliminary Results and Future Work}
\label{sec:results}

\begin{SCfigure}[][htbp]
    \includegraphics[scale=0.40]{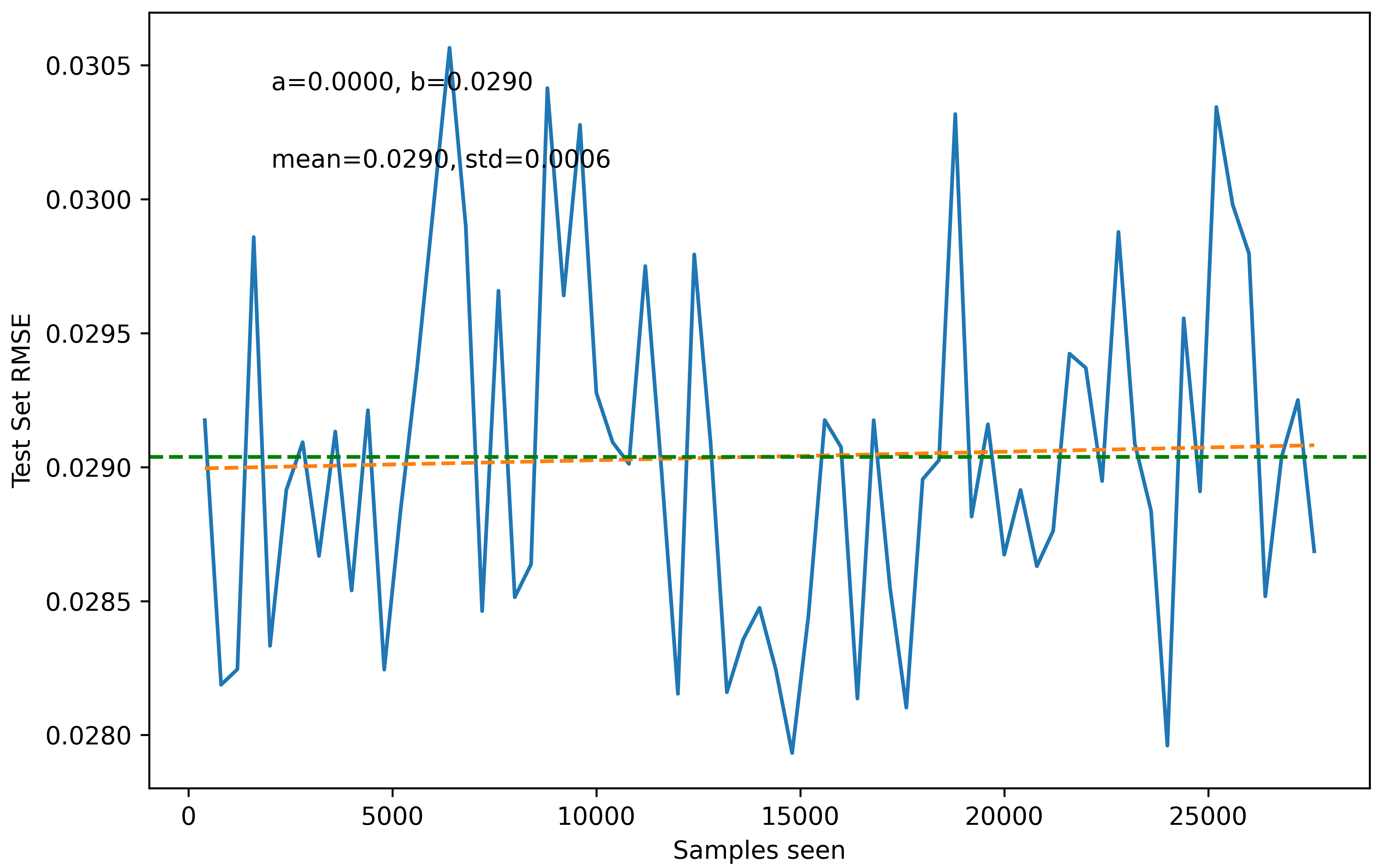}
    \caption{Test scores of the fully trained model on the test set, after 73000 iterations. 
    The green line shows the average error, and the orange is a linear regression.
    It illustrates the Root Mean Squared Error (RMSE) between the predicted and target LAI. 
    The reported mean RMSE is 0.03 and the standard deviation is 0.006 for a normalized LAI between -1 and 1 (for LAI ranging from 0 to 7). This represents a normalized RMSE LAI of 0.04 (for two standard deviations). 
    In an extreme scenario of $\text{LAI}=7$ (the maximum value in our dataset) this represents a RMSE of 0.3, which is well in line with acceptable errors for earth observation products, ranging in $\pm 0.5$ LAI \citep{FANG2012610}.}
    \label{fig_testing}
\end{SCfigure}

\begin{SCfigure}[][htbp]
    \includegraphics[width=0.5\textwidth]{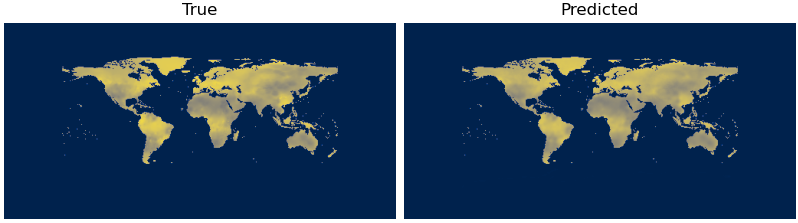}
    \caption{Comparison between a true one-month ahead observation (left column) and its prediction (right) from the test set. 
    The model provides an accurate reconstruction, with all major spatial patterns being well represented. Image in Cividis palette \citep{2017Nunez}, with zero LAI values in blue, and maximum ($\text{LAI}=7$) in yellow.}
    \label{fig_forecasting}
\end{SCfigure}

Preliminary results on the test set are illustrated in Figure~\ref{fig_testing}. The presented scores are in line with acceptable errors of earth observation products, ranging in $\pm 0.5$ LAI \citep{FANG2012610}. A qualitative result is also illustrated in Figure~\ref{fig_forecasting} showing the reconstruction capacities of the model. It depicts the ability of the model to predict both LAI variability and general anomalies with less in a day of training on a single NVIDIA A100 GPU.

The perspectives of this proposal could extend the time series range of the predictions, including a sensitivity study on the effects of feature lags on the predictions. An attribution study could be conducted to understand the contribution of the climate covariates on the LAI predictions. To this purpose, the global feature attribution and the local pixel attribution methods could be considered \citep{Molnar_2022}.

The subsequent work on eXplainable AI (XAI) feature attribution methods will attempt to score individual predictions considering each feature according to their contribution to the final prediction. The SHAP method (SHapley Additive exPlanations - \cite{2017LundbergLee}) could be used for local explanations while providing a global score for the model. Local pixel attribution methods, such as GradCAM \citep{2017Selvaraju}, would also be well suited to our proposed method to highlight relevant climate covariates with respect to the gradient of the model weights. Considering accurate LAI predictions, the proposed XAI methods could be able to untangle the factors driving global vegetation productivity and their relationships with global climate change.

\begin{ack}

This work was supported by the Climate Change AI Summer School (in-person cohort), and through usage of cloud computing credits provided by Denvr Dataworks. We acknowledge the support of the Natural Sciences and Engineering Research Council of Canada (NSERC), through its NSERC Alliance program (Application: ALLRP 566310-21) and the Canada Research Chair program (CRC-2019-00139). We also acknowledge the support of the McMaster University Centre for Climate Change Research Seed Fund. This work was developed in part with usage of the facilities of the Shared Hierarchical Academic Research Computing Network (SHARCNET), Calcul Québec, and the Digital Research Alliance of Canada through its Research Allocation Competition, under the 2023 Resources for Research Groups process “National carbon flux estimation system for forest ecosystems of Canada” (ID 4715).

\end{ack}

\small
\bibliographystyle{plainnat}
\bibliography{main.bib}

\end{document}